# Instruction and Solution Probabilities as Heuristics for Inductive Programming

Edward McDaid[1], Sarah McDaid[2]

Zoea Ltd (zoea.co.uk)

**Abstract**. Instruction subsets (ISs) are heuristics that can shrink the size of the inductive programming (IP) search space by tens of orders of magnitude. Here, we extend the IS approach by introducing instruction and solution probabilities as additional heuristics. Instruction probability reflects the expectation of an instruction occurring in a solution, based on the frequency of instruction occurrence in a large code sample. The solution probability for a partial or complete program is simply the product of all constituent instruction probabilities, including duplicates. We treat the minimum solution probabilities observed in code sample program units of different sizes as solution probability thresholds. These thresholds are used to prune the search space as partial solutions are constructed, thereby eliminating any branches containing unlikely combinations of instructions. The new approach has been evaluated using a large sample of human code. We tested two formulations of instruction probability: one based on instruction occurrence across the entire code sample and another that measured the distribution separately for each IS. Our results show that both variants produce substantial further reductions in the IP search space size of up to tens of orders of magnitude, depending on solution size. In combination with IS, reductions of over 100 orders of magnitude can be achieved. We also carried out cross-validation testing to show that the heuristics should work effectively with unseen code. The approach is described and the results and some ideas for future work are discussed.

## 1. Introduction

Inductive programming (IP) is a research area that aims to generate software directly from a specification, such as a set of test cases [1]. It is a field that has been active over several decades, resulting in a wide range of approaches, technologies and tools [2]. Throughout this period, it has also been beset by an apparently insurmountable problem.

Rice's theorem [3] implies that all non-trivial programs must be executed in order to determine their output for a given input. This means that any approach which produces working software according to a specification, must engage in some form of generate-and-test [4]. As a result, all inductive programming approaches are obliged search a space of candidate code solutions, in one way or another. The huge size of this search space has limited IP to the production of relatively small programs [5].

Conventional AI techniques for dealing with large search spaces are of limited use in the context of IP. For example, best first search based on the intermediate values produced by partial solutions will only work for some problems where an objective function can be defined and where the gradient is relatively consistent. Few real world programming problems have these characteristics. Indeed, there are many situations where no useful heuristic can be defined in terms of intermediate values or where no measure of distance correlates in any way with solution proximity. These factors have impeded the progress of IP and limited its adoption for decades.

In recent years the advent of large language models (LLMs) and their ability to generate code in response to prompts has attracted much interest [6,7]. However, in order to produce software that is guaranteed to work, LLM-based code generation must also engage in some type of generate-and-test. This aspect of their operation does not seem to garner a lot of attention.

As LLM-based tools are increasingly integrated into software development workflows, it is likely that they will have to utilise test cases and other specifications, in addition to text prompts, in order to guarantee the correctness of generated code. As a result, LLMs are likely to confront the same problems that have hampered IP for so long.

The search space size problem has been acknowledged in IP for as long as the field has existed [8]. However, the lack of any significant progress or even activity in addressing this issue is not encouraging. This suggests either that there is a dearth of ideas or that the problem is regarded as unsolvable.

[1]ORCID: 0000-0001-8684-0868

[2]ORCID: 0000-0001-7643-6722

**1.1 Zoea**

Zoea is an inductive programming system that has been in development for the last few years [9]. It was created specifically with the goal of addressing the IP search space problem and employs a number of strategies to this end.

Zoea generates code automatically from a set of test cases which can optionally include any number of intermediate values, in addition to inputs and outputs. Programs can also include other programs, allowing larger solutions to be assembled through composition. A visual programming version of Zoea is also available. Over the last couple of years Zoea has developed a family of novel and highly effective heuristics and hyper-heuristics focussed on the distribution of instructions in programs.

**1.2 Instruction Subsets**

Instruction subsets (ISs) are simple heuristics that aim to predict the set of instructions required to create a code solution for any problem [10]. This approach was motivated by the intuition that the search space for the relatively few instructions that are used in any given solution is much smaller than the search space for the complete instruction set.

Most high level programming languages include over 200 instructions, comprising operators, and core and standard library functions. This corresponds to a massive search space that grows rapidly with increasing program size. If it were possible somehow to predict the required set of instructions for a given problem then the search space would be exponentially smaller.

This intuition led to a study of the instruction co-occurrence patterns in a large sample of human code [10]. The study found that most possible pairs of instructions rarely or never co-occur while around 90% of program units (PUs) utilise 10 or less unique instructions and only 2% have 20 or more. Here, the term program unit refers to functions, subroutines and main code blocks.

The study also found that it was possible to cluster all of the instruction co-occurrence relationships contained in a huge code sample into a relatively small family of overlapping subsets of instructions. The number of subsets produced in this way depends on subset size but invariably is orders of magnitude smaller than the number of program units from which the subsets were derived.

The clustering process used to create subsets guarantees that each subset includes all of the instructions contained in the program units from which it was created. Conversely, this implies that the subset instructions define a search space that is capable of recreating all of these program units.

Clustering also means that each subset is capable of generating many other programs, corresponding to any intersection of its component PU subsets. This is a form of generalisation. Instruction subsets were found to generalise quickly with 10% of the code sample being enough to produce subsets that could generate 80% of unseen code.

In the previous study, instruction subset families with different subset sizes were created (10, 20, .. 100). The number of subsets required in each family depends on the subset size and these varied between around 6000 subsets at size 10 and 50 subsets at size 100.

Instruction subsets define search spaces that are up to tens of orders of magnitude (OOM) smaller than the size of the search space for the complete instruction set. The size of the reduction depends on the subset size and the size of the target program. The scale of the reduction is such that, that even with 100s or 1000s of subsets, the benefit is still enormous.

It is also interesting to note that even though the instruction subsets overlap, the proportion of redundant work quickly becomes vanishingly small as program size increases. This is because any redundant partial or complete solutions must be composed entirely from overlap instructions. This is a tiny fraction of combinations which exponentially approximates to zero with increasing solution size.

Every component within Zoea has the ability to utilise a defined subset of instructions. When we use a given IS within Zoea, it is as if only instructions that are members of that IS actually exist. All other instructions that are not present in IS are effectively irrelevant to the current problem. In operation, each instruction subset is used in turn to constrain the generation of candidate solutions in terms of which instructions are used. Instruction subsets can also be processed in parallel.

The success of the instruction subset approach inspired a number of follow-on studies. These resulted in the development of instruction digrams and type signature hyper-heuristics.

**1.3 Instruction Digrams**

Instruction digrams are a data flow model of instruction-instruction application that was also derived from a human code sample [11]. Each digram is simply a count of the number of times that the output of instruction A is used as an input to instruction B in all of the sample programs.

Instruction digrams are used to limit the ways in which multiple instructions can be assembled to form solutions by excluding patterns which do not occur in human code. When instruction digrams are applied to an existing instruction subset, the result is to eliminate an increasing number of instructions as data flow depth increases. This provides an additional search space size reduction of up to 5 orders of magnitude.

**1.4 Type Signatures**

The original instruction subset approach produced a single large family of instruction subsets for each subset size. Type signatures are hyper-heuristics that select one of many much smaller instruction subset families, based on the combination of data types present in test case or PU inputs and outputs [12].

These new instruction subset families are constructed using only PUs with matching type signatures. This avoids cases where a subset may be incapable of producing a solution since none of its instructions consume or produce a required data type. Type signature hyper-heuristics provide an additional search space reduction of up to 3 orders of magnitude.

## 2. Instruction Probability

The idea of using instruction probability as a search space heuristic has existed within Zoea in various forms for several years [13]. As with IS, this work is based on a simple intuition..

It has long been known that the instructions within a large body of human code have a Zipfian or power law distribution [14]. This means that a small number of instructions occur frequently while the majority occur rarely. Similar patterns are seen in the frequency of words in a large corpus of text.

The intuition behind the use of probability as a heuristic comes from the fact that most instructions in an instruction set have low probability, which implies that most possible combinations of instructions should include many rare instructions. However, this is not the case in human code.

Every PU is composed of one or more instructions and each instruction can occur any number of times within the same PU. In a code sample that contains many PUs, we can count the number of times a given instruction occurs across all PUs and also count the total number of instructions. In general terms, we can say that where instruction I has CI occurrences within a total number of instruction occurrences CT, then the instruction probability PI = CI / CT. Probability is a numeric value between 0 and 1.

We considered two formulations of instruction probability. Global instruction probability (PIG) is measured across the entire code sample. Calculation of PIG is straightforward and involves every instruction in all PUs. In addition, we can define instruction probability (PI) within a specific IS. PI for instruction I is calculated using the number of occurrences of I in all PUs covered by IS, divided by the total number of instructions in the same set of PUs. This means that a given instruction can have many different PI values - one for each IS of which it is a member. Each instruction only ever has a single value for PIG.

**2.1 Solution Probability**

We consider the occurrence of a single instruction in a complete or partial code solution as an event, with an associated probability. Therefore, the occurrence of two or more instructions in a solution corresponds to two or more events, again with associated probabilities.

Any ordering of these events is ignored and we effectively treat them as simultaneous. As a result, we assume that these events are independent, in the sense that the occurrence of one event does not alter the probability of any other event. We believe this to be largely true but also recognise that the reality is probably more subtle. Testing the validity of this assumption could be scope for future work.

The probability of two or more independent events is given by the product of their respective probabilities. This means that each solution or partial solution has an associated solution probability (PS) that is the product of the probabilities for all of their component instructions, including any duplicates:

$$PS(PU) = \prod_{I \in PU} PI(I)$$

Solution probability can be calculated using either PIG or PI. As noted above, each instruction can have different values for PI for each IS. This means that PS can also have different values for each IS that covers the solution. In addition, a single global solution probability (PSG) can be calculated using the PIG value for each instruction, again including any duplicates:

$$PSG(PU) = \prod_{I \in PU} PIG(I)$$

Solution probability only relates to the presence of instructions and does not reflect their order in the source code nor their logical structure. As such, it is intended to convey some indication of the relative likelihood that a given selection of instructions is used, as opposed to any other.

**2.2 Solution Probability as a Heuristic**

It should not be a surprise that most PUs in human code have relatively high solution probabilities. PSs are calculated using PIs and these instruction probabilities reflect the instruction distribution in the PUs from which the IS was composed. By definition, frequently occurring instructions have a high probability and there are many of them. Even though most instructions in an instruction set have low probabilities, there are relatively few infrequent instructions in human code, with low probabilities to drive down the product.

If a PU were to be composed entirely of low PI instructions then it would have a very low PS. The fact that PS is obtained through multiplication serves to amplify the range between the lowest and highest possible PSs for a given size.

It is theoretically possible for PUs with very low PSs to exist, but it is highly unlikely. Certainly, no such PU was encountered during the study. There are also very many possible PUs, that could exist in theory but have never been observed, with PSs that would be lower than any of those present in the code sample. This means that PS has the potential to be a useful heuristic during exploration of the IP search space.

When Zoea is solving a problem using a given IS, it assembles fragments of candidate solutions by incrementally adding instructions from the subset. If we already have the corresponding PIs or PIGs, pre-calculated from the code sample, it is an easy matter to calculate the PS or PSG for each such fragment. This can be used to allow us to avoid areas of the search space containing solution candidates with probabilities that are lower than some chosen threshold.

**2.3 Solution Probability Threshold**

For each solution size we can define a solution probability threshold (PST). This is the minimum PS that we are prepared to consider during search. By tracking the PS of each partial solution as they are assembled, the PST allows us to reject or postpone the addition of any instruction that would cross that threshold. This has the effect of avoiding potentially large areas of the search space.

A simple way to define PST for each solution size is to take the minimum PS value across all PUs of that size in the code sample. The code sample contains a very large number of PUs for each solution size so we can make the assumption that few if any future solutions would have a PS with a lower value. This choice of PST also means that - in principle - Zoea would be able to recreate all of the PUs in the code sample.

|  | **Global** | **Per IS** |
|---|---|---|
| Instruction count | CIG | CI |
| Instruction total | CTG | CT |
| Instruction probability | PIG | PI |
| Solution probability | PSG | PS |
| Solution probability threshold | PSGT | PST |

**Table 1.** Summary of acronyms

PIs and PSs are calculated with respect to a given IS. This implies that a different PST should similarly be defined for each solution size and for each IS. In contrast, a single global solution probability threshold (PSTG) can be defined in terms of PSG for each solution size. (Acronyms are summarised in Table 1.)

We can easily calculate the PS for each PU in the code sample for any given IS. Given the number of instructions in PU we can determine the minimum PS for each PU length (again for each IS). This measure reflects the values observed for real and complete PUs in the code sample.

For a given IS, we can also calculate the minimum and maximum possible PSs for each solution length. These are simply the lowest and highest PIs in the IS, raised to the power of length.

**3. Study**

The primary objective of the study was to investigate the usefulness of instruction probability, solution probability and solution probability threshold as heuristics for IP. We aimed to demonstrate this by measuring the size of the search space above the threshold for different solution lengths and calculating the relative reduction in search space size achieved with respect to a baseline search space size (IS size 10).

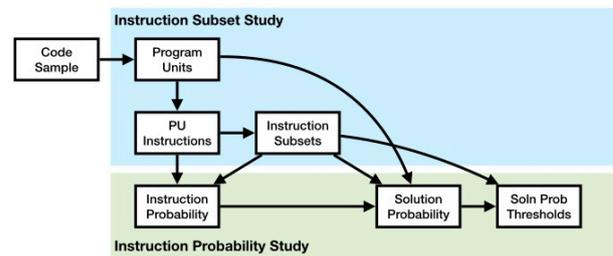

**Fig. 1.** Overview of study scope

**3.1 Scope**

Figure 1 provides an overview of the study scope, the data flow relationships between key study entities and the relationship to the IS study.

We reused the same code sample as the original instruction subsets study. For simplicity, the study is limited to instruction subsets of size 10, although this still corresponds to 90% of all PUs.

We reused the size 10 ISs from the original study. We also reused the IS search space size estimates for IS size 10. The number of occurrences of each instruction in each PU and across all PUs was already measured in previous studies.

**3.2 Data**

This study reused the same large sample of python code that was used in the original instruction subsets study and which is described in detail in that paper [10]. The sample comprises > 15.7 million lines of code from 998 large repositories.

This study utilised instruction subsets and a list of the instructions present in each PU. Both of these data products were already in existence.

### 3.3 Approach

The previous instruction subset study estimated the total size of the search space before and after the application of ISs. The use of estimation was unavoidable in previous studies due to the very large search space sizes involved.

In this study the reduction in search space size was such that precise measurement of the reduced search space size was feasible using exhaustive search and node counting. The use of instruction subsets dictated that we used the post-application IS size estimate as a baseline. The actual search space size measurements and the baseline allowed us to estimate the relative search space reduction for instruction probabilities.

The process mainly consisted of two similar strands of work: one for PIG and one for PI. In addition, cross-validation of thresholds against unseen PUs was also carried out.

### 3.4 PIG Process

The study process for PIG was as follows:
- Calculate PIG for each instruction I;
  - CIG(I) = count instruction I in all PUs;
  - CTG = count all instructions in all PUs;
  - PIG(I) = CIG(I) / CTG;
- For each PU with instructions PU[1..n]
  - PSG(PU) = PIG(PU[1]) * ... PIG(PU[n]);
- For each S = size(PU) = 1 .. 40:
  - PSGT(S) = min(PSG(PU)) where size(PU) = S;
  - Use S, PSGT and { PIG(I) } to perform an exhaustive search of all nodes, calculating PSG(node) and counting nodes where PSG(node) >= PSGT;
  - The reduced search space size corresponds to the node count;
  - The reduction ratio is calculated as the size of baseline divided by the reduced size.

### 3.5 PI Process

The study process for PI was as follows:
- For each IS where size(IS) = 10:
  - Calculate PI for each instruction I
    - CI(I) = count instruction I in all PUs where I ∈ PU and PU ⊆ IS and size(PU) = S;
    - CT = count all instructions in all PUs where PU ⊆ IS and size(PU) = S;
    - PI(I) = CI(I) / CT where I ∈ IS;
  - For each S = size(PU) = 1 .. 40:
    - Calculate PS for all PUs covered by IS where PU ⊆ IS and size(PU) = S
      - PS(PU) = PI(PU[1]) * ... PI(PU[n]);
    - PST = min(PS(PU)) for all PUs where PU ⊆ IS and size(PU) = S;
    - Use S, PST and { PI(I): I ∈ IS } to perform an exhaustive search of all nodes, calculating PS(node) and counting nodes where PS(node) >= PST;
    - The reduced search space size corresponds to the node count;
    - The reduction ratio is calculated as the size of baseline divided by the reduced size.

### 3.6 Validation Process

The validation process was as follows:
- Calculate PSG for all PUs using PIG;
- For training percentage PC1 = 0.0005% .. 50%:
  - Test percentage PC2 = 100% - PC1;
  - Split code sample PC1:PC2 -> L1, L2;
  - For each Size = 1 .. 40:
    - Thresh(Size) = min(PSG) in L1 where size(PU) = Size;
    - Result(PC1,Size) = percentage of PUs in L2 where PSG(PU) >= Thresh(Size) and size(PU) = Size

### 3.7 Challenge and Validation

A possible criticism, of both the heuristics and of the study, is that the instruction probabilities and thresholds might be overfitted to the solutions from which they were derived. This would mean that while the heuristics appear to work well, enabling the same solutions to be found quickly, they could provide less or no benefit in finding different ones.

This challenge can be addressed by cross-validation, which involves splitting the code sample into a training set and a test set. The training set would then be used to produce the thresholds, and these could then be tested using the test set.

Being able to create useful thresholds with a small training set is evidence that the approach also works for the PSs of unseen problems. Having a number of training sets of increasing size will show whether and how quickly this ability changes as the training set size increases.

For each training set percentage and each solution size, we can simply measure the percentage of PU solution probabilities in the test set that are higher than the corresponding threshold. This implies that that PU exists within the search space defined by the threshold, and would therefore be found in a search of that subspace.

### 3.8 Execution

Custom code was produced to orchestrate the process, measure the search space and capture the results. Measurements of the search space size were obtained for all PUs of size 1-40 and all ISs of size 10 using both PSG and PS. Cross-validation results were obtained for a range of training set size percentages from 0.0005% to 50%.

## 4. Results

### 4.1 Instruction Probability

Figure 2 shows the values for PIG (blue) and PI (red) for each instruction. Instructions (x-axis) are ranked by descending PIG value. There is a single PIG value for each instruction and one PI value for each IS. These values both have a power law distribution and range over approximately 4 OOMs.

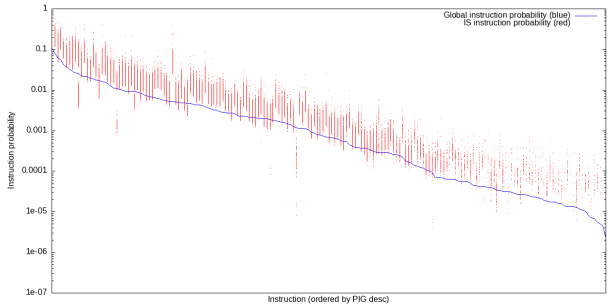

**Fig. 2.** PIG and PI instruction probabilities for each PU

Figure 3 shows PI distribution within all ISs. The x-axis corresponds to the instruction with the Nth highest PI value within each IS. It can be seen that there is a considerable variation in the range between the highest and lowest PIs across ISs.

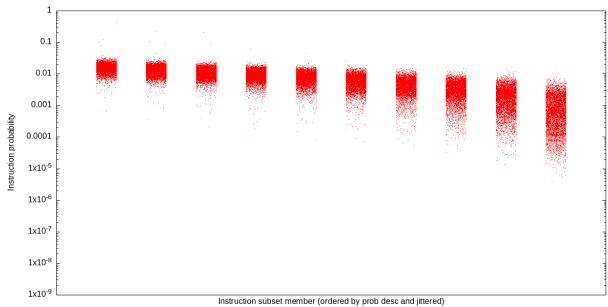

**Fig. 3.** PI instruction probabilities across each IS

### 4.2 Solution Probability

Figure 4 shows the PSG (blue) and PS (red) solution probabilities for each PU, in increasing PSG order. There are relatively few PUs with extremely low probabilities while most have relatively high probability. As solution probability decreases PSG and PS become less correlated.

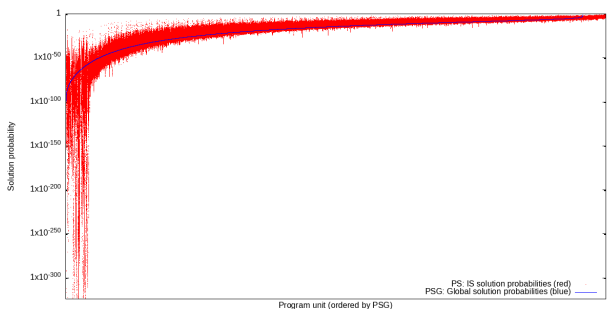

**Fig. 4.** PSG and PS solution probabilities for each PU

Figure 5 shows the ranges of PS and possible IS solution probabilities for a typical IS (size 10 #19). A similar plot was produced for each of the thousands of size 10 ISs. The x-axis corresponds to PU size. The green lines show the minimum, median and maximum PS for all PUs that are covered by this IS, for each size. The red lines show the minimum, median and maximum possible solution probabilities that could be created with this IS, for the given solution size.

It can be seen there is a large and rapidly widening gap between the probabilities that are observed in code sample PUs and the probabilities of possible solutions that could be created.

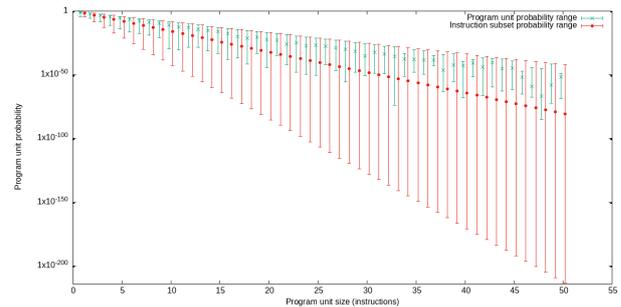

**Fig. 5.** PU and IS solution probability range example

Figure 6 is similar to figure 5 but instead shows data for all PUs and all ISs on a single chart. The green and blue points show the maximum and minimum possible solution probabilities for each IS. The red points show the solution probabilities for all ISs. Again these are shown by increasing solution size and jittered for visibility. This shows that PU probability tends towards the high end of the range with an increasing skew towards much lower possible probabilities - particularly with larger solutions.

In other words, human code seems to have a significantly higher solution probability than random code of the same size.

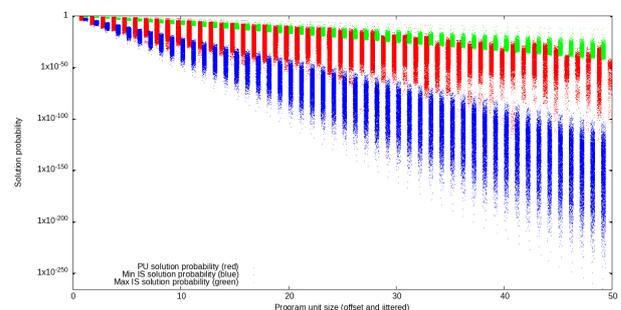

**Fig. 6.** All PU and solution probability ranges

### 4.3 Search Space Size Reduction

Figure 7 shows the relative search space size reduction achieved using PSG (blue) and PS (red), for different solution sizes. The x-axis is jittered such that PIG and PI are displayed side by side. It is clear that both perform very similarly in giving very large reductions of tens of orders of magnitude. This trend also seems to be slightly increasing, particularly with larger solution sizes. This is further evidenced in Figure 8.

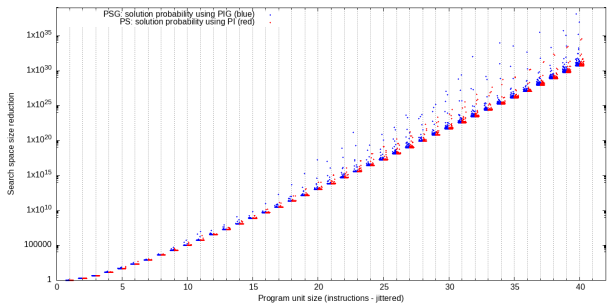

**Fig. 7.** Relative search space size reduction

Figure 8 shows the actual search space sizes measured for each PU using PSG (blue) and PS (red). Again, there is very little difference between the sizes achieved with the two approaches. What is clear is that the sizes achieved are tens of OOMs smaller than the IS baseline and at size 40 around 80 OOMs smaller than the search space size without heuristics. If the trend continues to size 50 then the reduction would be over 100 OOMs.

Another interesting feature is that the rate of increase in search space size seems to reduce with increasing solution size. Given this is a log plot the flattening out really corresponds to linear growth.

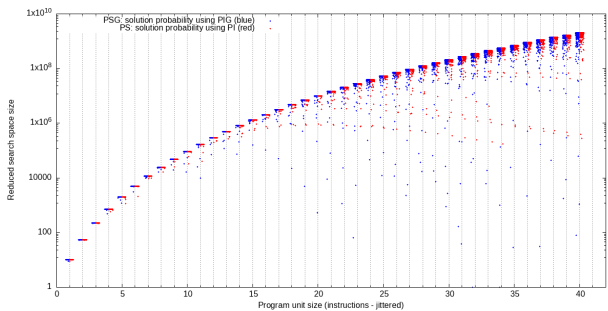

**Fig. 8.** Actual search space sizes for PSG and PS

Figure 9 illustrates the overall scale of the search space reduction. It shows the same plot as Figure 8 together with the total size of the space for each solution size without any heuristics (black) and with IS only (green).

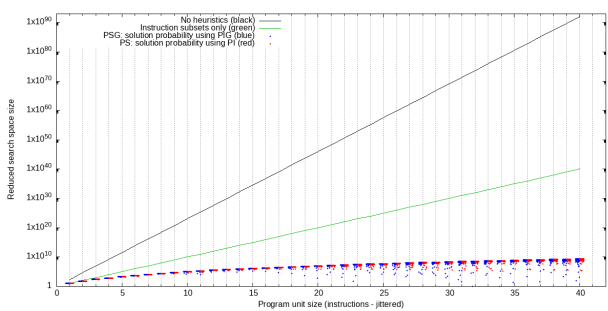

**Fig. 9.** Overall reductions with IS plus PSG or PS

Figure 9 also shows that the current heuristics represent a large increase in the size of programs that can be generated. On this plot a search space size of around $10^{20}$ represents the upper limit of what is currently achievable using Zoea. (It doesn't include the reductions for digrams or type signatures.) For IS only, this corresponds to around 20 instructions while with solution probability, even 40 instructions is comfortably within range. Indeed, if the trend continues then significantly larger sizes would also be well within reach. Determination of the actual size limit is important future work.

### 4.4 PSTG Versus PST

The results for PST and PTSG in Figures 7 and 8 appear very similar. In addition, there is no significant difference between the corresponding sums of their search space sizes for all PUs.

This implies that IS does not impact instruction or solution probabilities in any practical way and that PIG and its derivative metrics should be used, since PIG is computationally more straightforward than PI. There is nothing wrong with PI and its derivatives and they work just as well as their PIG counterparts. There is simply no advantage in using PI to justify the additional complexity and work.

### 4.5 Cross-Validation

Figure 10 shows the results of cross-validation testing. Different percentages (0.0005% to 50%) of the code sample were used as a training set to define thresholds for each solution size (1-40). For each set of thresholds, the remainder of the code sample was used as a test set to determine the percentage of PUs of each size that were within the search space defined by the thresholds.

It can be seen in Figure 10 that threshold coverage in the test set rises very quickly with increasing training set size. Only 5% of the code sample is required to achieve 99% coverage of unseen PUs, in most cases. This is evidence that the heuristics are not overfitted to the code sample and suggests that they are also likely to work well with unseen future code.

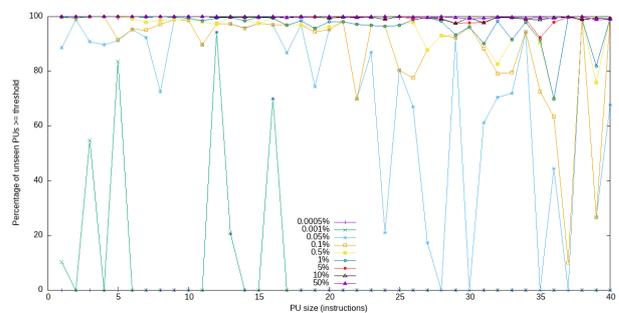

**Fig. 10.** Cross-validation with a range of training set percentages

## 5. Discussion and Future Work

The use of instruction and solution probabilities (ISP) as heuristics delivers very significant reductions in the IP search space size. This has been measured against the - already greatly reduced - search space sizes obtained using instruction subsets. For example, with solution size 40, IS reduces the search space size by around 50 OOMs while ISP further reduces this by another 30 OOMs. For larger solutions the total reductions are even greater. At size 50, the total reduction with IS and ISP combined would be over 100 OOMs.

The search space size obtained with IS 10 was used as the baseline. This was necessary in order to be able to measure the marginal reduction in search space size provided by ISP, while also using instruction subsets. It also means that there can be no doubt that the benefits of these heuristics are indeed cumulative - as opposed to being double-counted or interdependent in some way.

It is interesting that the rate of increase in search space size seems to attenuate with larger solution sizes (Fig 8). Based on this, it may be that the search space size using ISP never gets much higher than $10^{10}$ - even for the largest programs. Other heuristics - particularly instruction digrams - will reduce this even further but these sorts of sizes no longer present any real challenge, even in terms of exhaustive search. Nevertheless, it would be useful to fully understand the reason for this attenuation.

The concept of solution probability introduced in this paper is defined solely in terms of instruction probability. In addition, it may also be possible to create further heuristics by leveraging the probabilities of other solution characteristics, such as topology, complexity, branching factor or data flow depth. Whether these potential heuristics should be combined with instruction probability, or treated separately, is an open question. There is considerable scope for future work here together with some potentially useful heuristics.

A key assumption in this study is that the probabilities of solutions to future problems will be within the ranges defined by the PSTs derived from the code sample. This is partly mitigated by the use of a large and diverse sample which is the result of many different coders. While it seems likely that at least many such future problems will have similar probabilities, the risk that some may not remains unquantified. Without a supply of unforeseen problems that can be used for testing, verification of this assumption can only come through extended application of the heuristics to new problems in production.

The study did not investigate the use of instruction probabilities with either instruction digrams or test signatures. Both of these approaches work in ways that should not be impacted by instruction probabilities, or vice versa. Therefore, there is every reason to believe that the latest results are also cumulative with respect to these other heuristics as well. This assumption will be verified in the future through testing on the completely integrated stack.

The PSTs used in the study were treated as static. PST could also be modified automatically as part of search. For example, it might be increased at first to consider only higher probabilities before being expanded to cover progressively lower probabilities. This would have the effect of starting with a very small search space in an attempt to find only high probability solutions. This strategy would be successful in a significant percentage of cases. If a solution is not found then a larger search space can be explored using lower PST values, and so on. This type of approach is similar in principle to iterative deepening depth-first search [15].

In the study, PUs from the code sample were used as the direct source of the PST values. Higher or lower PST values could also be used. Higher PST values might be used where the lowest PS is seen as an outlier, causing the search space to be larger than necessary. Conversely, lower PST values would enlarge the search space, allowing a wider range of solutions to be produced. Selecting the best values for PSTs is a key factor in optimising the approach and is an important focus for future work.

In operation, PSTs could also be updated dynamically over time. For example, runtime performance metrics could be used to build a model of PST impact on search duration. This could then be used to adapt search space size to the allotted time and resource budgets for a task.

Whatever values are used for PST there is always a small possibility that some problems would require a still lower value in order to produce a solution for a given IS. It may be useful to quantify the target performance as a percentage of problems for which a solution is obtained. Having said that, it should also be remembered that there are often many functionally equivalent solutions for any problem and some of these may be found using different ISs, with lower PSs for equivalent solutions.

For simplicity, the study was limited to IS size 10. There is every reason to believe that the approach would also work for IS size 20 and above. While it is considered unlikely, if it turns out that the approach only works for size 10, it would still be useful, as this accounts for 90% of PUs. In any event, this will be verified and quantified through testing on the fully integrated stack.

The study treats the probabilities associated with the presence of two or more instructions as independent. It is unknown whether this is truly the case and if not the scale of any such impact is also unknown. This would be interesting future work.

This work does not necessarily imply that human-like code has some intrinsic merit over non-human-like code, or vice versa. However, this would be an interesting topic for further study. What is certain is that humans are frequently able to create code solutions and that these occupy a part of the configuration space associated with high solution probabilities. Our new approach takes advantage of the fact that this is an efficient and often fruitful region in which to search.

It is envisaged that the current work will lead to significant reductions in the technical resources required for Zoea to produce solutions. This will allow much larger solutions to be produced in less time with the same resources. It should also make it feasible to run Zoea locally on commodity hardware or even on mobile devices.

As with instruction subsets, a generalisation of the ISP approach may have applicability in other areas beyond IP. For example, a derivative of the approach could be used with any kind of production system, such as rules or grammars, where probabilities can be associated with the occurrence of different solution elements, based on some corpus or training set. While probability is already widely employed in approaches such as probabilistic grammars and fuzzy logic, these existing uses tend to be generative, in the sense of directing reasoning towards solutions. In contrast, an approach derived from ISP would also enable unlikely candidate solutions to be excluded - together with potentially large parts of the search space that they occupy. This represents an additional and complementary benefit in these problem areas.

Studies such as the one described here can lead to more significant technological advances and performance improvements than years of tedious and expensive engineering. Indeed, no degree of tuning or economically achievable level of scaling would match the benefits of IS or ISP in the foreseeable future. Also, simply waiting for computers to get faster is not a useful strategy, even in the long term. For these reasons, similar studies are expected to continue as long as promising ideas remain unexplored.

A number of additional candidate heuristics have been identified and some of these are outlined in this and previous papers. Studies of these further heuristics have not yet commenced.

Some - or possibly all - of the process of identifying and deriving similar heuristics from code sample data could be automated. All of these heuristics share a common and relatively simple conceptual model of code. Similarly, the processes used and the functional elements that enable their creation have a high degree of commonality. All of these heuristics are basically an answer to the question: what piece of information separates sample code from the largest possible subset of non-sample code? There are many possible answers to this question so any degree of automation in the process of finding and creating heuristics would be very useful.

## 5.1 Ethics

The data extracted from the code sample and used in this study was limited to lists of instruction identifiers. These lists of instructions are not unique to any given program, developer, repository or project, but rather are very common and widely duplicated. As such, they convey no useful information about the code they came from and also cannot represent anyones intellectual property. Therefore, we believe that the extraction and use of this information is entirely ethical and does not violate the rights of coders or anyone else.

## 6. Conclusions

We have introduced instruction and solution probabilities, and associated thresholds as additional heuristics for inductive programming. These were evaluated in a study using IS size 10 as a baseline. Two variants of instruction probabilities were tested and found to perform nearly identically. Both gave significant further reductions in search space size of up to tens of orders of magnitude, relative to the baseline. This represents an overall reduction in combination with IS of around 80 orders of magnitude at solution size 40. Reductions of over 100 orders of magnitude may be achieved for larger solutions. The resulting search space sizes were measured directly and are amenable to exhaustive search even with modest resources. Interestingly, the rate of increase in reduced search space size appears to slow with larger solutions. Cross-validation testing shows that the heuristics work well with unseen code. The results were discussed and some opportunities for further work identified.

## Acknowledgements

This work was supported entirely by Zoea Ltd. (https://zoea.co.uk/) Zoea is a registered trademark of Zoea Ltd. All other trademarks are the property of their respective owners. Copyright © Zoea Ltd. 2025. All rights reserved.